\documentclass[jpcm]{iopart}

\usepackage{epsf,citesort}
\usepackage{graphicx,citesort,amssymb}

\begin{document}

\title{Theory and simulation of gelation, arrest and yielding in attracting
colloids}

\author{M. E. Cates$^1$, M. Fuchs$^2$, K. Kroy$^3$, 
W. C. K. Poon$^1$, A. M. Puertas$^4$}
\address{$^1$ School of Physics, JCMB Kings Buildings, The University of
Edinburgh, Mayfield Road, Edinburgh EH9 3JZ, United Kingdom}
\address{$^2$ Fachbereich Physik, Universitaet Konstanz, D-78457 Konstanz, Germany}
\address{$^3$ Hahn--Meitner Institut, Glienicker Str.~100, 14109
Berlin, Germany}
\address{$^4$ Group of Complex Fluid Physics, Department of Applied Physics, University of Almeira, 04120 Almeira, Spain}

\begin{abstract}
We present some recent theory and simulation results addressing the
phenomena of colloidal gelation at both high and low volume fractions,
in the presence of short-range attractive interactions. We discuss the ability of
mode-coupling theory and its adaptations to address situations with
strong heterogeneity in density and/or dynamics. We include a discussion
of the effect of attractions on the shear-thinning and yield behaviour under
flow.
\end{abstract}

\maketitle

\section{Introduction}
Recent studies of mode-coupling theory (MCT) have predicted various
kinetic arrest scenarios for colloids with short-range attractions
\cite{bergenholtz-fuchs:99,fabbian-etal:99,dawson-etal:2001}.  The
behaviour at the arrest transition has been analysed in considerable
detail as a function of the volume fraction $\phi$ of the colloids,
their attraction range $\delta $ (in units of the particle radius $a$)
and the attraction $\varepsilon$ (in units of $k_BT$)
\cite{goetze-sperl:2003,sperl:2004}. This approach, which builds on a
successful theory of the glass transition for hard spheres, suggests
that the underlying `ideal glass transition', found within MCT, could
provide a universal mechanism for homogeneous gelation in which the
arrested state is viewed as an attraction-driven glass
\cite{poonthis,pham-etal:2002}.

Here we do not attempt to mediate between those who support and those
who oppose the basic philosophy of MCT (see \cite{paris,capri} for
discussions). Instead, in Sections \ref{phase}-\ref{shear}, we outline
three areas of recent work that examine the ability of MCT to deal with
heterogeneous systems. In Section \ref{phase} we consider attractive
colloids at low density, and consider kinetic arrest by
routes that combine aspects of irreversible aggregation, gelation, and
phase separation.  The phenomenology that we describe takes MCT as a
reliable theory of arrest at high densities only, and attempts to
extend the same picture to the much lower colloid volume fractions at
which gelation actually occurs in systems with strong short-range
attractions. The resulting gels are very heterogeneous and our
approach is to preserve what we can of the MCT theory while allowing
for heterogeneity in density at scales either shorter or longer than
those at which we apply the MCT. In the first case (shorter scale
heterogeneity) one has a picture involving the MCT-like arrest of
clusters; in the second (longer scale) one has a phase-separation
driven morphology in which one phase arrests into a dense gel or
attractive glass.

In Section \ref{sim} we turn to gelation at higher particle densities,
where unmodified MCT can reasonably hope to succeed, at least in the
absence of phase separation. Here we carefully analyse simulation data
for attractive colloids, suppressing phase separation by including a
weak, long-range repulsion. We find that MCT gives an excellent
account of averaged dynamical quantities but that this disguises an
underlying dynamics which is much richer, and which in fact shows many
hallmarks of dynamical heterogeneity (DH): populations of fast and
slow-moving particles coexist. This heterogeneity appears to be
closely related to static heterogeneity of density which is present
for attractive colloids even at quite high volume fractions. This form
of DH may thus be different from that found in hard spheres
\cite{crocker} where the density fluctuations are smaller.  The new
type of DH may be enhanced by the long-range repulsion we have added,
but even if that turns out to be true, the results are of strong
interest. They show firstly that when experimental data agrees with
MCT this does not exclude the possibility of DH; likewise that
observation of DH does not exclude the possibility that MCT remains
predictive for averaged quantities such as the dynamic structure
factor on the fluid side of the transition. 

In Section \ref{shear} we describe some related recent work on
colloids under shear, and give a preliminary account of calculations that address the effect of short range attractions on the rheological behaviour. This offers nontrivial predictions, in particular for the variation of yield stress
and shear thinning behaviour with the range of attraction
$\delta$. In Section \ref{conclude} we offer some
concluding remarks.

\section{Gelation at low density}
\label{phase}
\subsection{MCT, phase separation, and aggregation}
For attractive colloids, MCT appears to be very useful at predicting
averaged dynamical quantities like the dynamic structure factor
$S(q,t)$, so long as the volume fraction $\phi$ of colloid is high
enough \cite{puertas-fuchs-cates:2002,pham-etal:2002,langmuir}. In the following we
refer to the resulting spatially homogenous gel as type I. In
practice, for moderately short relative range of attraction $\delta
\gtrsim 10^{-2}$  
a metastable gas-liquid phase
separation \cite{poon-haw:97} interferes with this kinetic arrest
scenario. On quenching to create a gel, phase separation is likely to
intervene for all colloid volume fractions below the intersection of
the binodal with the MCT arrest line.
 
The `hidden' binodal, when present, dominates over the much slower
crystallization route to the thermodynamic ground state, so that it is
liquid-gas separation not crystallization that can interfere more
strongly with glass formation. Moreover, due to the \emph{kinetic} (as
opposed to thermodynamic) mechanism underlying the arrest transition,
this potentially complex interplay sensitively depends on the quench
rate relative to the typical time scale for phase separation. To a
first approximation, the latter is given by the Smoluchowski time,
which is the time scale for doublets to form by colloidal collisions. 
Depending on whether the quench is fast or slow
with respect to this natural time scale, we expect transient gels to
develop along one of two different routes, in the following referred
to as type II and type III, respectively. Type-II gels are homogeneous
on short scales but strongly heterogeneous at the mesoscale: they
result when the characteristic coarsening textures produced by phase
separation get `frozen in' during the coarsening process, as a result
of an MCT-like arrest of one of the two phases
\cite{sedgwick-etal:tbp}. Type-III gels are, in contrast, assemblies
of long-lived nonequilibrium structures locally resembling those
obtained from irreversible cluster aggregation \cite{vicsek:92} and
thus heterogeneous also on short scales. To distinguish from fully
irreversible aggregates, in which the bonds formed are permanent, the
type III process is sometimes called `weak gelation'.

During the process leading to type-II gels, the system remains in (or
sufficiently close to) local equilibrium, so that gelation by this
route represents a relatively straightforward combination of phase
separation and MCT. However, additional concepts are needed for the
type-III scenario. Some of us have recently proposed a schematic
description (called cluster-MCT or CMCT) \cite{kroy-cates-poon:2004}
for weak gelation in such suspensions, in terms of an effective theory
for a fluid of `renormalised particles'. These represent
coarse-grained clusters and their interactions are analysed, in an
MCT--like fashion, to predict the onset of global kinetic arrest.  The
underlying view of gelation as a double ergodicity breaking (on the
monomer scale \emph{and} on the cluster scale) seems to be supported
by recent numerical simulations \cite{delgado-etal:2003}.

\subsection{Schematic nonequilibrium phase diagram}
\begin{figure}
\begin{flushright}
\includegraphics[width=0.8\textwidth]{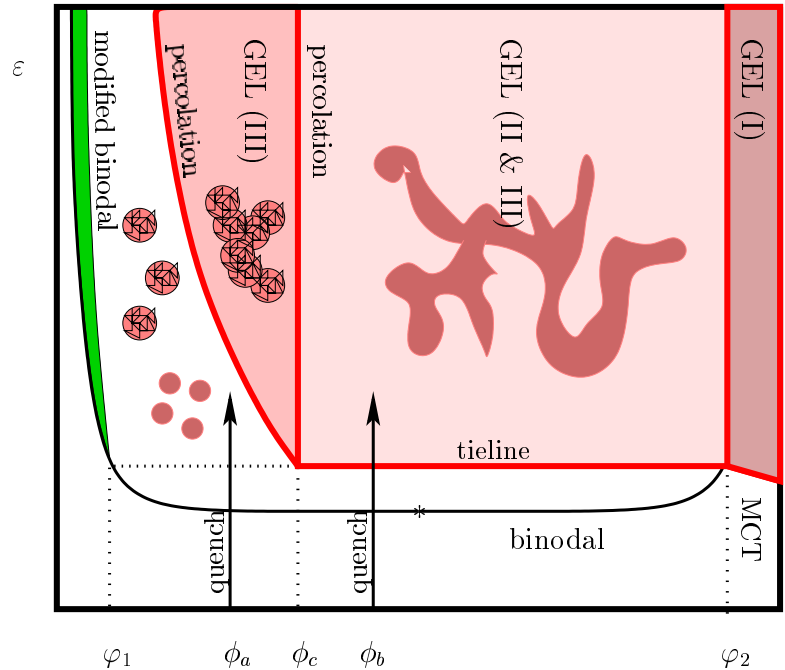}
\end{flushright}
\caption{Schematic ($\phi-\varepsilon$) cut through the phase diagram of
adhesive hard spheres. The interaction of gas-liquid phase separation
with MCT kinetic arrest gives rise to a variety of metastable gel
phases. At high colloid volume fractions $\phi$, where MCT can be
applied directly, homogeneous gels form (region I). At lower $\phi$
phase separation may create macroscopically heterogeneous gels (region
II) and (homogeneous) gel beads if the quench is slow, while more
tenuous gels (III) and (ramified) clusters consisting
of non--equilibrium particle aggregates result from rapid
quenches.}\label{fig:phases}
\end{figure}

A schematic phase diagram of the different nonequilibrium behaviours
predicted for hard sphere colloids with short-range attractions is
provided in Fig.~\ref{fig:phases}.

\emph{Slow quench scenario:} We concentrate first on this conceptually
simple case, which results in type-II gels. Consider a slow quench
along the quench path indicated by the arrow at volume fraction
$\phi_a$. Upon crossing the (metastable) liquid-gas binodal, the fluid
starts to decompose into a minority phase consisting of colloidal
`liquid' domains (rich in particles), and a majority `gas' phase
consisting of domains where colloids are scarce. These domains undergo
slow coarsening, while their compositions continue to evolve along the
binodal in response to the ongoing quench. During this process, the
two phases both remain in local coexistence on the binodal and move
apart until the compositions $\varphi_1$ and $\varphi_2$ are
reached. Note that, although the quench is slow, we assume that
macroscopic phase separation is even slower, so that the morphology is
one of mesoscopic domains. Once the colloid volume fraction within the
minority phase reaches $\varphi_2$ (see Fig.~\ref{fig:phases}), which
is where the binodal cuts the MCT line, the denser domains undergo
kinetic arrest: the characteristic time scales for further structural
evolution suddenly increase dramatically. From the lever rule, the
volume fraction of space occupied by the resulting amorphous solid
phase is $\Phi(\phi) = (\phi-\varphi_1)/(\varphi_2-\varphi_1)$.

Depending on the mesoscale domain structure that has accompanied the
preceding phase separation, this volume fraction $\Phi$ may be more or
less than the value $\Phi_c$ required for the minority phase to
percolate. Only if it does percolate do we have a macroscopic solid or
gel: note that gelation, in the sense of a finite elastic modulus,
requires {\em percolation of an arrested phase}, or equivalently
arrest of a percolated one. In particular, existence in the system of
a percolating network of bonds, each of which is transient, is not
sufficient to create a modulus.

In terms of the colloid volume fraction $\phi$, we obtain a
percolation threshold $\phi_c$ defined by $\Phi(\phi_c) = \Phi_c$. For
a volume fraction $\phi_b > \phi_c$ as depicted in the figure, a
connected gel, of finite modulus, is the predicted result of a
quench. By varying the colloid volume fraction and the quench
kinetics, different gel morphologies should be realizable in a
relatively well-controlled manner. Indeed, corresponding recipes are
routinely applied in the industrial processing of colloidal and
polymeric gels for optimizing rheological properties (see e.g.\
\cite{dickinson:92,owen-jones:98}).  Upon decreasing $\phi$, the gel
gets more teneous and eventually, for $\phi <\phi_c$ (e.g., $\phi =
\phi_a$ in Fig.~\ref{fig:phases}) this disintegrates into a fluid of
slightly sticky gel beads. (At still lower densities, there
is a slight shift in the binodal curve because the coexisting dense phase
cannot be denser than $\phi_2$.)

For moderately deep quenches, the mutual attraction between the
colloids, though strong enough to solidify dense colloidal drops
(forming the aforementioned beads), is insufficient to permanently
bind beads that come into contact.  This is because the beads are now
made of colloidal glass, and cannot adapt their shapes to allow
coalescence on the Smoluchowski timescale, even though this would
reduce their surface energy significantly. Two such beads of radius
$R$, treated as effective particles, have a far smaller relative range
of attraction than the primary particles ($\delta_{\rm eff} \sim
\delta/R$), which for roughened surfaces may be only partly outweighed
by its increased depth $\varepsilon_{\rm eff}$ \footnote{Note that for
ideal spheres $\varepsilon_{\rm eff}$ would typically increase
linearly with size and thus over--compensate the effect of the
reduction in the effective range $\delta_{\rm eff}$. However, the bead
surfaces are expected to be roughened due to particle deposition in
the ongoing quench, so that estimating the effective attraction is a
subtle task left for future work.}. For deeper quenches, the residual
interaction between beads may become strong enough to cause a second
round of phase separation and/or gelation at the bead level; the whole
argument can be iterated to describe that case.

Ignoring this last effect for the moment, the gel region in the phase
diagram is bounded by a section of liquid-gas tieline emanating from
the intersection point (at $\varphi_2$) of the binodal with the MCT
arrest line, and a line parallel to the quench route, emanating from
this tieline at the volume fraction
$\phi=\phi_c\equiv\varphi_1+\Phi_c(\varphi_2-\varphi_1)$ corresponding
to percolation. The transition across the first boundary is
`temperature--driven' (or more accurately $\varepsilon$-driven) and
governed by the MCT arrest of the dense phase; quenching by this
route, some features of the MCT transition (which applies directly
only for $\phi\ge\varphi_2$) are expected to be detectable dynamically
at much lower concentrations \cite{kroy-cates-poon:2004}. In contrast,
the `pressure--driven' (or more accurately, $\phi$-driven) transition
across the remaining section of the gel boundary should fall into the
percolation universality class \cite{lironis-heermann-binder:90}.  The
possible secondary gelation of arrested beads at deep quenches
complicates this picture somewhat; we have neglected it in the figure
and do not pursue this here.

\emph{Rapid quench scenario:} Rapid quenches are ones in which the
system finds itself under conditions of strong, quasi-irreversible
aggregation, so that bonds created in collisions between particles are
very long lived. In this case one expects at low volume fractions a
nontrivial episode of structure formation, akin to irreversible
diffusion-limited aggregation, on a timescale fast compared to other
processes. Morphologically, rapid quenches should differ
from the slow quenches described above, with a more ramified local
structure and a correspondingly larger region in which an arrested
phase percolates; the gel phase thus extends to lower volume fractions
than in a slow quench. 

The resulting type-III gels are conceptually distinct from the rest,
and may have some intriguing properties. Their analysis is more
difficult, as can be appreciated by considering a deep rapid quench at
low volume fractions. This is the case considered in the CMCT theory
of Ref.\cite{kroy-cates-poon:2004}. The basic idea of CMCT is to allow
for strong, quasi-irreversible bonding at short length scales by
applying MCT, not to primary particles, but to fractal aggregates
created by such bonding. This is quite similar to the idea already
introduced above, of iterating MCT to describe the possible gelation
and arrest of dense beads. However, the calculation of the effective
particle parameters is even more subtle and, in particular, allows for an
entropic decrease in the effective attraction between clusters. This
reflects the multiplicity of internal bonds at which two clusters,
having joined to form a larger one, can now be broken apart to recover
two clusters of similar size and shape to the original pair
\cite{kroy-cates-poon:2004}.  Another potentially important feature is
the ability of CMCT to allow for a buildup of long-range repulsions
due, for example, to a very slight Coulomb repulsion that may be
present in many colloids, even in organic solvents
\cite{yethiraj-blaaderen:2003}. This buildup also lowers the effective
attraction strength as the clusters get larger.

The effect of the scale-dependent effective bond strength, combined
with a scale-dependent range, means that in CMCT there is a tendency,
as aggregation proceeds, to move away from the attraction-driven
arrest scenario and towards a more conventional repulsion-driven
glass. Setting aside phase separation effects (for now), the key
issue in gelation is then whether the effective attractions become
small while the volume fraction of clusters is still fairly low, or
whether, by the time they stop aggregating, the clusters are dense
enough to be arrested anyway by repulsive caging. In the first case
one predicts a semi-ergodic phase comprising a fluid of clusters
\cite{segre-etal:2001}; in the second, an arrested cluster phase which
is nonergodic at all scales and thus a gel
\cite{kroy-cates-poon:2004}.

For type-III gels the interplay with phase separation is quite
complicated; roughly speaking it follows the lines already developed
above for slow quenches, but with CMCT-like arrest replacing the
standard MCT arrest throughout the discussion. However the phase
separation itself is also perturbed by what is happening at the
cluster scale, which is a further complication. (In mitigation, for
many systems the dense phase will be dense enough for the CMCT and MCT
predictions to nearly coincide through much of the phase diagram
anyway.) Within the CMCT picture there can be two distinct forms of
cluster phase: one in which there is no tendency to macroscopic phase
separation but clusters stop growing due to the buildup of repulsion
(Coulombic or entropic); another in which phases have separated but
the dense arrested phase does not percolate
\cite{kroy-cates-poon:2004}. The latter includes the gel beads
described previously as a limiting case \cite{sedgwick-etal:tbp}.

CMCT is based on a simplifying assumption that the timescale for
internal reconstruction of clusters is slow compared to the timescale
for realizing a state of repulsion-driven arrest at larger
scales. This may be safe for Coulombic stabilization, but needs
careful further test, against both experiment and simulation, in the
case where the bond strength is effectively reduced by the entropy
associated with bond breaking internal to a cluster \cite{Sciortsim}.

\subsection{Combining MCT with phase separation: further remarks}
\label{longrange}
By translating knowledge of the interaction potential into a combined
topographic map of the MCT transition surface and the gas-liquid
coexistence, we were able to present above and in
\cite{kroy-cates-poon:2004} (see also \cite{sedgwick-etal:tbp}) a
guide for disentangling some of the experimentally observed complex
phenomenology of attracting colloids.

We now take a somewhat broader perspective in which, on top of the
hard sphere repulsions, the pair potential can be represented as the
sum of a strong short-range attraction and an additional weak
long-range interaction that may either be attractive of repulsive.
Numerous examples ---from microemulsions, through block copolymers, to
supercooled water--- show that such competing interactions can give
rise to complex phase behavior. Even the adhesive hard-sphere system
(AHSS) of Baxter \cite{baxter:68} is now known to have two competing
crystalline phases, two glasses, and a nontrivial re-entry between
them. This stems from a competition between the hard-core repulsion
and the short range attraction \cite{sciortino:2002,poonthis}. Adding
further long-range interactions can certainly complicate this further.

\begin{figure}
\begin{flushright}
\includegraphics[width=0.8\textwidth]{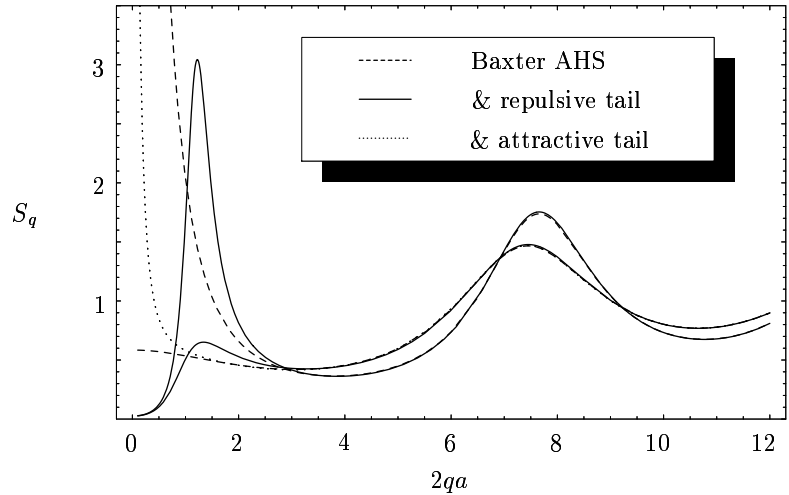}
\end{flushright}
\caption{Static structure factor $S_q$ for the AHSS \cite{baxter:68}
with volume fraction $\phi=0.3$ and adhesiveness parameter $\tau=0.08$
and $\tau=0.2$, respectively (the former being closer to the spinodal
and showing the more pronounced oscillations), calculated by
a random phase approximation. Three cases of
competing interactions are distinguished: The bare AHSS (dashed
lines), the AHSS with an additional repulsive Yukawa barrier of height
$0.1$ and decay length 2.5 (solid lines), and with an attractive tail
of the same range but with an amplitude 
$-0.004$ (dotted line, only for $\tau = 0.2$ since $\tau = 0.08$ lies
within the spinodal region).}\label{fig:sq}
\end{figure}

Insofar as it destabilizes the equilibrium liquid, the origin of this
complexity is manifest in the static structure factor $S_q$
\cite{sear-gelbart:99}; this is also the input for MCT
calculations. In Fig.~\ref{fig:sq}, $S_q$ is shown for an AHSS with
additional long range interaction. The low--$q$ tail of the structure
factor shows the characteristic upturn found near any spinodal. A
sharper upturn is seen (well before the spinodal is reached) if a very
weak but longer ranged \emph{attraction} is added (dotted). In
contrast, adding a \emph{repulsive} barrier can energetically prohibit
macroscopic phase separation as evidenced by the suppression of the
spinodal divergence and the shifting of the new relative maximum in
$S_q$ to finite wave vectors $q=q_c$ (solid lines). Near the former
spinodal, the new peak grows in height without bound, thus heralding
microphase separation, in which the homogeneous fluid decomposes into
finite, self-limiting domains of liquid-like and gas-like
character. The domain size ($\simeq q_c^{-1}$) is controlled by (and
roughly scales as) the range of the extra repulsion.

To predict the non-equilibrium behaviour in the presence of the competing
interactions, note first that thermodynamics and kinetic arrest are
sensitive to different features of the pair potential. Gas-liquid
phase separation can be induced by a weak long-ranged attraction, as
familiar from van der Waals' theory; and similarly, even finite
repulsions may cause (Wigner-)crystallization if sufficiently
long-ranged. Attraction-driven arrest, on the other hand, relies on
kinetic trapping of a thermodynamically stable fluid by short-ranged
forces ($\delta\ll1$), and according to MCT the arrest is triggered by
the high Fourier modes $qa\simeq \delta^{-1}\gg 1$, as encoded in the
direct correlation function $c_q\propto 1-S_q^{-1}$
\cite{bergenholtz-fuchs:99}. This contrasts with the crowding of the
nearest-neighbour shell visible in the first peak of $S_q$ near
$qa\simeq \pi$, which leads to the ``cage effect'' for repulsive hard
sphere glasses.

Recently there has been much debate whether the long range repulsion
itself \cite{klein-etal:94}, or the frustration resulting from the
competing interactions \cite{westfahl-schmalian-wolynes:2001}, may be
a new driving mechanism for the formation of more exotic arrested
states. Indeed, apart from the repulsive (and attractive) hard-core
glasses familiar in colloids, MCT allows for another transition
scenario triggered by long range repulsions that give rise to a dilute
arrested state, which was interpreted as a \emph{Wigner glass}
\cite{bosse-wilke:98}. Even liquid-crystalline domains of asymmetric
particles have been predicted to undergo kinetic arrest
\cite{letz-schilling-latz:2000}. Along somewhat different routes
similar conclusions were reached about so-called \emph{stripe glasses}
in microemulsions and related systems
\cite{westfahl-schmalian-wolynes:2001}.  Several of the mentioned
mechanisms (attractive/repulsive `hard-core' interactions of arrested
spinodal textures; a glass transition triggered by a microphase peak
in $S_q$; the Wigner glass) may be relevant to the possible gel phases
of Coulomb-stabilised cluster fluids
\cite{yethiraj-blaaderen:2003,segre-etal:2001}.  More generally we
therefore expect that the interplay of MCT with microphase separation
could lead to a phenomenology at least as rich and interesting as the
one elaborated above for the case of bulk phase separation.

\if{
(more on repulsion stabilizing bead and cluster phase)

\emph{Summary}

\begin{table}\label{tab:gl}
\begin{tabular}{|c|cc|cc|}
\hline
 quench rate  & \multicolumn{2}{c|}{rapid} & \multicolumn{2}{c|}{slow}   \\
 quench depth  & shallow  & deep  & shallow
 & deep  \\ 
\hline
micro--scale & liquid--gas  & ICA clusters & repeated l.--g.
 & glass, gel  \\
macro--scale & drops $\leftrightarrow$  bicont.\ &  clusters $\leftrightarrow$ gel & 
nested separ.\ &  beads $\leftrightarrow$  gel \\
\hline
\end{tabular}
\caption{Different scenarios for MCT--arrested gas--liqiud phase
separation and micro--phase separation. Different quench rates with
respect to the rate of phase separation lead to different
micro--structures of the arrested phase.  Structures formed upon
ordinary phase separation are transient while in presence of a
competing long--range repulsion, drop-- and cluster--phases can be
thermodynamically stable. The arrow $\leftrightarrow$ indicates
percolation transitions in the first case, while in the second case
macroscopic arrest of drops, clusters or beads can additionally be
caused by the long--range forces (``Wigner crystal/glass'').}
\end{table}
}
\fi


\section{Simulation of dense attractive colloids}
\label{sim}
We now turn to our second theme, which is to test the predictions of
MCT for dense attractive colloids, in a regime of concentration and interactions where
these predictions could be reasonably expected to work. Thus we can
bypass the various complications connected with phase separation and
aggregation that arise at low density, as were considered in Section
\ref{phase}.

\subsection{Simulation details}

Newtonian dynamics simulations were performed to test the theoretical
predictions on the gel transition in a system whose pair potential
mimics the depletion attraction found in colloid polymer mixtures
\cite{poonthis}. One thousand polydisperse particles were considered
in the canonical ensemble. The core-core repulsion between particles
is given by $V_{sc}(r)\:=\:k_BT\left(r/a_{12}\right)^{-36}$, where
$a_{12}=a_1+a_2$, with $a_1$ and $a_2$ the radii of the particles. A
flat distribution of radii was used to prevent crystallisation, with a
(half-)width equal to one tenth of mean radius, $\Delta=0.1 a$. The
interaction between the colloidal spheres is given by the
Asakura-Oosawa interaction potential, which considers the depletion of
ideal polymers \cite{asakura54}, corrected to take into account the
polydispersity of the colloids \cite{mendez00}. The total potential
($V_{\mbox{tot}}=V_{AO}+V_{sc}$) was corrected close to $a_{12}$ to
ensure that the minimum of the total potential is at $a_{12}$
\cite{puertas03}.

A long-range repulsive barrier was added to the interaction potential
in order to prevent liquid-gas separation at high attraction
strength. When this barrier is allowed for, the potential is
attractive at short distances, and slightly repulsive at longer
distances. The range of the attraction is given by the size of the
polymers, $\xi$, and the strength at contact, $r=a_{12}$, is
proportional to the polymer volume fraction $\phi_p$. The height of
the repulsive barrier is $1 k_BT$ whereas arrest occurs for
attraction strengths above $\sim 7k_BT$ \cite{puertas03}. While the
remarks of Section \ref{longrange} serve as a warning that this kind
of barrier could influence the results, our aim here is merely to
avoid the liquid-gas separation that would otherwise complicate
matters considerably. The barrier is thus chosen as small as is
consistent with achieving this.

In our simulations, lengths are measured in units of the mean particle
radius, $a$, and time in units of $\sqrt{4a^2/3v^2}$, with $v$ the
thermal velocity, which is set to $\sqrt{4/3}$. The equations of
motion were integrated using the velocity-Verlet algorithm, with a
time step of $0.0025$. The volume fraction of colloids is fixed at
$\phi=0.40$; $\xi = 0.1a$; and the attraction strength is parameterised
by the polymer volume fraction, $\phi_p$.

\subsection{Suppression of phase separation}

Separation of the system into two phases of different density was
monitored by a demixing order parameter, $\psi_4$. The system is
divided in $4^3$ boxes of equal size, and the deviation of the density
in every box with respect to the average is measured by this
parameter; an homogeneous system shows $\psi_4$ close to zero and a
phase-separated system has a much bigger value. With this parameter,
the isochore $\phi=0.40$ was first studied for the system without a
long range barrier, in order to see any effect gelation might have on
liquid-gas separation (and, potentially, vice versa).

\begin{figure}
\begin{flushright}
\includegraphics[width=0.8\textwidth]{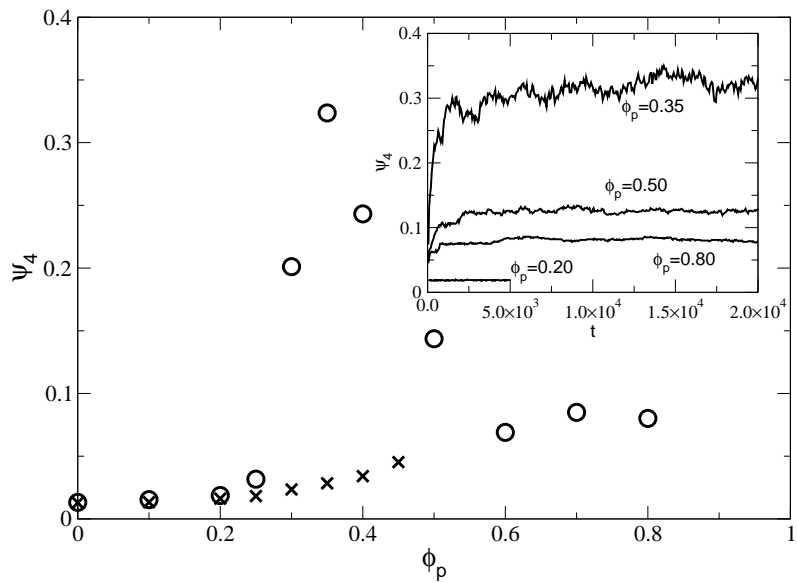}
\end{flushright}
\caption{\label{fig1} Long time limit of the demixing order parameter,
$\psi_4$ as a function of the polymer volume fraction,
$\phi_p$. Inset: Evolution of $\psi_4$ with time for different
$\phi_p$ as labelled.}
\end{figure}

The inset to \fref{fig1} shows the temporal evolution of $\psi_4$ for
different polymer fractions. At low $\phi_p$, the system is
homogeneous, i.e., the attraction strength is too low to induce
liquid-gas separation, and for $\phi_p=0.35$, the system demixes, as
shown by the increase of $\psi_4$. At even higher attraction
strengths, the system does not phase separate into denser and more
dilute phases, as expected, but becomes again more homogeneous. The
long time limit of the separation parameter, which is plotted in \fref{fig1}
as a function of $\phi_p$, captures this scenario. The
liquid-gas transition is found to take place at $\phi_p=0.30$, in
agreement with previous simulations \cite{dijkstra99}. The $\psi_4$
parameter does not increase monotonically, as expected from
equilibrium thermodynamics, but reaches a maximum and decreases. A
plateau at $\psi_4 \approx 0.08$ is observed above $\phi_p=0.60$,
indicating a quasi-homogeneous system. We may thus conclude that some
mechanism must be present that hinders liquid gas separation at high
attraction strength and finally prevents it. This mechanism is
gelation, i.e., attraction-driven arrest.

\subsection{Results for averaged quantities}
To analyze the arrest mechanism, the long-range repulsive barrier
was restored to the interaction potential, thus forbidding liquid-gas
separation and allowing the gelation to be probed in its own
right. With the barrier, the system is macroscopically homogeneous at
all polymer fractions studied, although it presents detectable voids
and `tunnels' when viewed from certain angles, and 
also shows a low-angle peak in the structure factor at $qa \sim 1$
\cite{puertas03}. However this is always lower than the primary near-neighbour peak at $qa\sim \pi$.

\begin{figure}
\begin{flushright}
\includegraphics[width=0.8\textwidth]{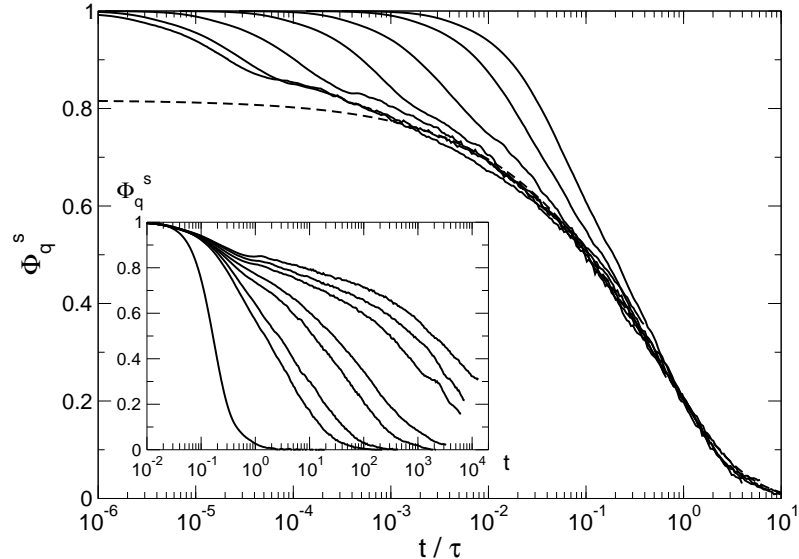}
\end{flushright}
\caption{\label{fig2} Density-density correlation functions for
increasing polymer fraction, rescaled at long times. From left to
right $\phi_p=0.425,0.42,0.415,0.41,0.40.0.39$ and $0.375$, and
Kohlrausch fitting to the $\alpha$-decay (dashed curve). Inset:
Unscaled correlation functions for the same states, from right to
left. The hard-sphere case, $\phi_p=0$ is added (left curve).}
\end{figure}

The dynamics of the system close to gelation is studied by means of
the self part of the density-density correlation function,
i.e. $\Phi_q^s(t)=\langle \exp \left\{i {\bf q} .\left[{\bf
r}_j(t)-{\bf r}_j(0)\right] \right\} \rangle$, where the brackets
indicate averaging over particle $j$ and time origin, and ${\bf q}$ is
the wavevector. \Fref{fig2} shows the density correlator for
increasing polymer fraction for $qa=6.9$, the second (non-microphase) 
peak in the
structure factor. The two step decay in the correlation functions is
similar to the behaviour of states approaching the glass transition in
the Lennard-Jones system (LJS) or in hard spheres (HS)
\cite{kob95,kob95a}. Furthermore, the decay from the plateau can be
rescaled at long times for all the states presented
(with deviations for the state closest to the transition, as seen also in other simulations \cite{kob95,kob95a})
and the master
decay can be fitted using the Kohlrausch form, $\Phi_q^s(t)=A_q
\exp(-(t/\tau_K)^{\beta})$, a signature of MCT-like nonergodicity
transitions. Similar scalings are obtained at all the wavevectors
studied (not presented here).

Using the predictions from MCT, the early decay from the plateau is
correctly described by the von Schweidler law:
$\Phi_q^s(t)=f_q^s-h_q^{(1)} (t/\tau)^b+h_q^{(2)}
(t/\tau)^{2b}+O(t^{3b})$, where $f_q^s$ is the non-ergodicity
parameter and $h_q^{(1)}$ and $h_q^{(2)}$ are amplitudes. All three of
these are state-independent (specifically, independent of $\phi_p$), whereas $\tau$ is a time scale which
carries state-dependent information, increasing as the glass
transition is approached. The von Schweidler expression correctly
describes our correlation functions for all states and wavevectors,
and the results from the fittings show that $f_q^s$ in this
attractive case is much bigger than the nonergodicity parameters found
in HS or LJS. This fact indicates that the localization length is much
shorter than in those cases, showing that the driving mechanism for
this transition is the formation of long-lived bonds between
particles. Moreover, the von Schweidler exponent, $b$, from the
fittings also differs from the HS or LJS values, yielding $b=0.37$
\cite{puertas03}, in agreement with the predictions from MCT
\cite{dawson-etal:2001}.

\begin{figure}
\begin{flushright}
\includegraphics[width=0.8\textwidth]{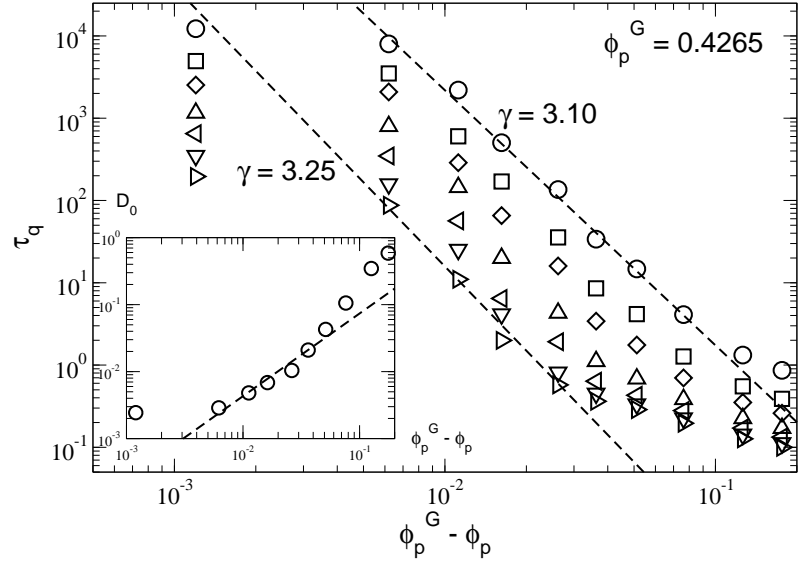}
\end{flushright}
\caption{\label{fig3} Wavevector dependent time scale, $\tau_q$
vs. $\phi_p^G-\phi_p$ for different wavevectors. From top to bottom
$qa=3.9, 6.9, 9.9, 15, 20, 25, 30$. The lines are power law
fittings. Inset: Self diffusion coefficient, $D_s$,
vs. $\phi_p^G-\phi_p$ and power law fitting.}
\end{figure}

For the time scales $\tau_q$ of the $\alpha$ decay, defined by
$\Phi_q^s(\tau_q)=f_q^s/e$, MCT predicts a power law divergence, with
an exponent $\gamma$ related to $b$. \Fref{fig3} presents the time
scales at different wavevectors on logarithmic axes. The power-law
divergence is very clear, and the exponents agree with the value
expected from $b$, which is $\gamma=3.1$. The gel point, $\phi_p^G$, also
fitted in the analysis, is found at $\phi_p^G=0.4265$ for this volume
fraction, $\phi=0.40$. It is interesting to note that the
competition between gelation and liquid-gas separation at this polymer
fraction was already apparent in our study of the system without the
long range barrier, although gelation fully impedes phase separation
only well above this value of $\phi_p$.

MCT also predicts a power-law decay, with the same exponent $\gamma$,
for the self-diffusion coefficient, $D_0$. This coefficient is
determined from the long time behaviour of the mean squared
displacement, $\langle \delta r^2 \rangle=6 D_0 t$. Fixing the gel
point to the value reported above, the fitted exponent in this case is
$\gamma=1.23$, quite different from the value obtained above for the
divergence of the time scale. Such differences are obtained in the
analysis of other model systems, LJS or HS, and have been attributed
to the presence of dynamical heterogeneities in the system. However in
our work the difference between the exponents determined from the time
scale and the diffusion coefficient is larger than in these other
cases. This suggests a possible stronger role for dynamic
heterogeneities in the presence of short-range attractions.

\subsection{Dynamical heterogeneity}
\begin{figure}
\begin{flushright}
\includegraphics[width=0.8\textwidth]{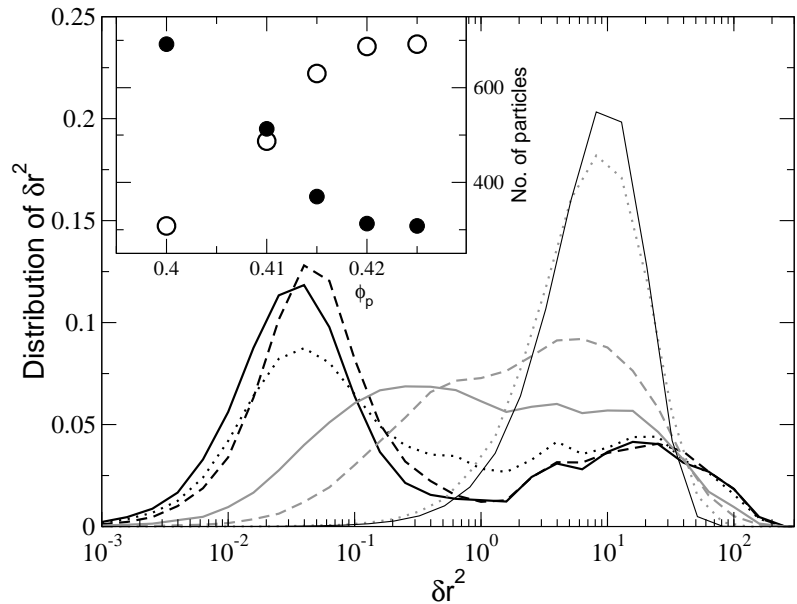}
\end{flushright}
\caption{\label{fig4} Distribution of squared displacement for
$\phi_p=0.30$ (gray dotted line), $\phi_p=0.40$ (gray dashed line),
$\phi_p=0.41$ (gray solid line), $\phi_p=0.415$ (black dotted line),
$\phi_p=0.42$ (black dashed line), and $\phi_p=0.425$ (black solid
line). In all cases $\langle \delta r^2\rangle=10 a^2$. The thin black
line is the theoretical distribution for a system of non-interacting
Brownian particles. Inset: Number of slow particles (open circles) and
fast particles (closed circles) in the system as a function of the
polymer volume fraction.}
\end{figure}

The dynamical heterogeneities of a state can be studied by analysing
the distribution of the squared displacement of particles measured
between some arbitrary time $t=0$ and a later time $t=t^*$. A
homogeneous fluid should present a single peaked distribution, its
width depending on $t^*$ and $D_0$. \Fref{fig4} presents this
distribution for different states approaching the gel transition,
where $t^*$ has been chosen so that $\langle \delta r^2(t^*)
\rangle=10 a^2$ for all states. At low polymer fractions, the system
is indeed homogeneous and the distribution of squared displacements
agrees with the theoretical expectation for a system of Brownian
particles. However, at higher attraction strengths, the system becomes
more and more heterogeneous and the distribution more bimodal.

For polymer volume fractions above $\phi_p=0.41$ two peaks are clearly
observed in the distribution, showing one population of particles that
move much less than a particle radius and another population of very
mobile particles. (These of course combine such that the average
squared displacement is $\langle \delta r^2 \rangle=10 a^2$, since
this was used to define $t^*$.) In fact, these two populations of
particles can be distinguished at all times. The exchange between them
is detectable, but very slow, allowing an analysis of the system to
proceed as if it were composed of two entirely distinct populations:
`fast' particles and `slow' ones \cite{puertas04}.

The nontrivial structure of the incipient gel (with voids and tunnels) makes
possible the existence of a subset of particles fully integrated into
a percolating gel-like structure, each with a large number of neighbours,
while also allowing another subset of particles to be present at the
surface of the same structure and thus to have only a few
neighbours. The slow particles form a stiff structure which is very slow to
relax, whereas the fast particles surround this structure, with a high
mobility. Since there is a relatively large amount of free space in
the system, the fast particles move freely and not in clusters or in
string like motions. The fast particles attach from time to time to
the incipient gel structure at surface sites but generally depart again before
getting trapped. Preferential sites for this transient adsorbtion of
fast particles are detected in pockets of the incipient gel formed by
the slow ones, where they can establish more bonds with slow particles
\cite{puertas04}. 

\subsection{Implications for MCT}

It is interesting to point out that the average behaviour of the
system is correctly described by MCT \cite{puertas03}, both in the
universal properties and the specific predictions for the attraction
driven glass transition. The main exceptions to this involve the
non-gaussian parameter (not shown; see \cite{puertas04}) and the
discrepancy between $\gamma$ as measured from diffusion and from
structural relaxation. However, the detailed dynamics, involving two
concurrent populations of particles, signifies very strong dynamical
heterogeneities which are not really consistent with the approximation
scheme lying behind MCT calculations. The strong DH observed in our
incipient colloidal gels is related, it appears, to structural
heterogeneities which have no direct analogue in the glass transition driven by
repulsion \cite{crocker}. At lower densities, these heterogeneities
are even more important; in that region one expects to require
significant modifications to MCT, as developed and discussed in
Section \ref{phase}.

\section{MCT under shear}
\if{We note in passing that the framework of MCT can be extended to
address the relationship between stress and strain-rate in a system
undergoing shear \cite{fuchsprl,fuchsrest,capri}. This includes a route,
currently somewhat technical, to relate the yield stress of an
arrested phase to its static structure factor $S_q$ as determined in
an unstrained state. The response is always shear thinning. A
schematic extension of this approach, allowing for a stress-dependent
MCT vertex, admits the possibility of shear thickening states, which
are often observed experimentally in dense repulsive colloids
\cite{colin}. (Some of these states might become fully jammed -- a
fluid that ceases to flow because of application of a stress.) We
mention these developments here because attractive interactions may
have strong and interesting effects on all these calculations. This
avenue is left for future work.
\label{shear}}\fi

The framework of MCT can be extended to
address the relationship between stress and strain-rate in a system
undergoing shear \cite{fuchsprl,fuchsrest,capri}; this includes a way to relate the yield stress of an
arrested phase to its static structure factor $S_q$ as determined in
an unstrained state. 
Via $S_{q}$, the effects of  attractive interactions on the
nonlinear stress response can be incorporated. 
The response is always shear thinning. A
schematic extension of this approach, allowing for a stress-dependent
MCT vertex, admits the possibility of shear thickening states, which
are often observed experimentally in dense repulsive colloids
\cite{colin}. Although we expect attractive 
interactions also to have
interesting effects on shear-thickening,
we focus below on the shear-thinning case.
 
The approach of \cite{fuchsprl,capri}
predicts a finite yield stress $\sigma^+$ at the arrest 
transition. 
For gels of attracting colloids, 
the scaling of $\sigma^+$ with the width of the attraction is found
below by 
combining a virial-type analysis of the AHSS model \cite{bergenh00}
with the model of 
`isotropically sheared hard spheres' (ISHSM) \cite{fuchsprl,fuchsrest}. 
(This model isotropises advection in its effect on density
fluctuations.)
The
resulting scaling can be compared to the value
$\sigma^+=0.75 k_BT/a^3$ found at the glass
transition of  the ISHSM \cite{fuchsrest}.

The ISHSM derives the nonlinear flow curves (viz., stress $\sigma$
versus strain rate 
$\dot \gamma$) from the competition between the slowing down of the
structural dynamics  
captured in the classical MCT, and the speeding up of fluctuations caused
by shear advection. 
The latter mechanism has been intensively studied in the context
of fluctuations in 
sheared systems close to criticality \cite{Onuki}, and close to
ordering or microphase 
transitions \cite{CatesMilner}. It causes a time dependence of the wavevector
of an arbitrary fluctuation
\begin{equation}\label{advek}
{\bf q}(t)= ( q_x , q_y + q_x \dot \gamma t, q_z )\; ,
\end{equation}
where the shear is along the $x$-direction with $v_x= {\dot
  \gamma}y$; 
$t$ here is the time since the start of shearing (or the birth
of a fluctuation). If a fluctuation
  initially has 
a wavelength $\sim 1/q$, at a later time $t$ its wavelength
$\sim |{\bf q}(t)|^{-1} = 1/q(t)$ will be smaller, leading to decay of
  the fluctuation 
caused by fast, small-scale particle rearrangements.

The scaling of the yield stress $\sigma^+$ with the attraction range $\delta$ can,
for small $\delta$, 
be estimated from a virial expansion of $S_{q}$ for the AHSS: one finds
$1- S^{-1}_{q} \to 6 A \phi \delta\, \sin(2 q a)/(q a)$, where
$A=\exp{\varepsilon}-1$.  
Inserting this expression into the ISHSM \cite{fuchsrest}, one can
take the limit $\phi\to 0$ and
$A\to\infty$ with $\Gamma_v = 6 \phi A^2\delta/\pi^2$ 
held fixed. This leads to the longitudinal
memory kernel:
\begin{equation}\label{memor}
m_{\tilde{q}}(t) \to \frac{\Gamma_v}{2\tilde{q}^2} \; \int^{\tilde{k}_>}\!\!\!
 d^3\tilde{k}\; 
\left( \frac{{\bf \tilde{q}}\cdot {\bf \tilde{k}}}{\tilde q \tilde k}
 \right)^2\; 
\cos\left({\frac{\tilde{k}(t)-\tilde k}{\delta/2}}\right)   \; \Phi_{\tilde k}(t)
\Phi_{|{\bf \tilde q}-{\bf \tilde k}|} \; ,
\end{equation}
while the corresponding expression for the steady state shear stress becomes:
\begin{equation}\label{spannung}
\sigma \to {\dot \gamma}\; \frac{k_B T}{a^3}\; \frac{\phi
  \Gamma_v}{5\delta^2} \; 
\int_0^\infty \!\!\!dt\; 
 \int_0^{\tilde{k}_>}\!\!\!
 d\tilde{k}\; \tilde k^2\;
\cos\left({\frac{\tilde{k}(t)-\tilde k}{\delta/2}}\right)   \; \Phi^2_{\tilde k}(t) \; .
\end{equation}
Here, the limit $\delta \to0$ applies, and the
rescaled wavectors $\tilde q= q a \delta$ and a 
cutoff $\tilde k_>$ 
were introduced; choosing $\tilde k_>\approx3.68$ maps
the AHSS virial results onto the 
corresponding results of an attractive square well system with range
$\delta$ \cite{bergenh00}. 

In the approach of Refs.\cite{fuchsprl,fuchsrest},
these equations are closed via an MCT-like relationship
for the normalized 
density correlation functions 
$\Phi_{\tilde q}(t)$, whence
we obtain predictions for the
nonlinear rheology close 
to the arrest transition (at $\Gamma_v^c=1.42$) of attracting colloids 
with small $\delta$ and low $\phi$. 
Although the low--$\phi$
approximation appears drastic, if shear is switched off in Eq.\ref{memor}, 
the results for the
arrest line at small $\delta$ agree qualitatively
with those of Ref.\cite{dawson-etal:2001} up to rather large $\phi$ 
\cite{bergenh00,langmuir}.
(Thus the approximations made are faithful to the 
standard MCT
analysis, though they neglect the physics of heterogeneity discussed
in Section \ref{phase}.)
With shear present, we can now analyse the altered behaviour of 
$\Phi_{\tilde q}(t)$ at the onset of arrest and then compute the
limiting stress at low shear rate by inserting the resulting asymptotics
into Eq.\ref{spannung} \cite{fuchsrest}. This gives a
finite yield stress $\sigma^+$ just within the glass, which falls abruptly
to zero on entering the fluid.

Asymptotically the $\Phi_{\tilde q}(t)$  obey a `yielding scaling law', $\Phi_{\tilde q}(t) \to
\Phi^+_{\tilde q}(\hat t )$ with $\hat t = t / \tau(\dot\gamma)$ and $\tau(\dot\gamma)$ a
shear-rate dependent characteristic time. Also one 
finds the limiting closure relation
\cite{fuchsrest}: 
\begin{equation}\label{skalenges} 
\Phi^+_{\tilde q}(\hat t ) = m^+_{\tilde q}(\hat t ) - \frac{d}{d \hat t} \;
\int_0^{\hat t} \!\!\!dt' \; m^+_{\tilde q}(\hat t-t')\;
\Phi^+_{\tilde q}(\hat t ) \; , 
\end{equation}
with $m^+_{\tilde q}(\hat t)$ obtained from Eq.\ref{memor} using
the limiting form $\Phi^+_{\tilde q}(\hat t)$. 
The yield stress $\sigma^+$ then follows from inserting
$\Phi^+_{\tilde q}(\hat t)$ into 
Eq.\ref{spannung}. 

The expressions for the memory kernel and stress, Eqs.\ref{memor},\ref{spannung}, contain a 
rapidly oscillating term $\cos(2({\tilde{k}(t)-\tilde
k)/\delta})$, which arises from 
interference of the particle density fluctuations
within the narrow region of attraction. Studying this factor illuminates the role of shear. Without shear, 
constructive interference holds, and the factor is unity;
the memory kernels describe as usual bond-formation owing to
attraction, and the correlator stays arrested at its glass value,
$\Phi^+_{\tilde q}(\hat t) = \Phi^+_{\tilde q}(0) = 
f^c_{\tilde q}$. 
Under shear, the advection of wavevectors (giving
$\tilde{k}(t)-\tilde k= (k_x k_y/k) \dot \gamma t +\ldots$) produces
rapid oscillations in this 
term when $\delta$ is small. The interference is destroyed, causing
a fast decay of the 
memory functions. The time scale $\tau(\dot \gamma)$ needs to
be found self-consistently 
from Eqs.\ref{memor} and \ref{skalenges}; the preceding
argument shows that it scales with shear-rate and
attraction range 
as  $\tau(\dot \gamma) = c_{\rm a} \delta \, / |\dot \gamma|$ where
$c_{\rm a}$ is of order unity.
In contrast, for repulsive interactions in the ISHSM, we found
\cite{fuchsprl,fuchsrest}
$\tau(\dot \gamma) = c_{\rm r} / |\dot \gamma|$ with $c_{\rm r}$ again of order
unity.
Integrating Eq.\ref{spannung}, one obtains (with further
constants $c'_{a,r}$) the
scaling expressions 
$\sigma^+_{\rm AHSS} = c'_{\rm a}\,G_{\rm a} \, \delta$ and
$\sigma^+_{\rm ISHSM} = c'_{\rm r}\,G_{\rm r}$ for the yield stresses at the two transitions.  Here $G_{\rm a,r}$ are the corresponding
shear moduli which, within MCT, are predicted to acquire finite
values on arrest. Their scalings have been discussed elsewhere
\cite{bergenholtz-fuchs:99,dawson-etal:2001,langmuir,bergenh00,Rama}; for AHSS one has 
$G_{\rm a} \sim kTa^{-3}\delta^{-2}$ whereas
$G_{\rm r} \simeq kTa^{-3}$ holds for hard spheres. 

The final scaling result for AHSS is thus $\sigma^+\simeq G_{\rm a}\delta \simeq kT a^{-3}\delta^{-1}$. This dependence of the yield stress on the attraction
range leads to interesting 
scenarios in systems where both attraction and repulsion-driven
arrest can be
observed \cite{pham-etal:2002,Bartsch}, and where at high
enough concentrations the local structural dynamics should dominate the
rheological behaviour.   
Interpreting the ratio of yield stress to the elastic modulus as a yield strain
$u_y = \sigma^+/G$, the scaling dependence of $\sigma^+$
on $\delta$ can easily be understood. The yield strain of
a (high density) 
colloidal gel is of order the relative range of the attraction $\delta$: the solid is
shear-melted as soon as  
particle bonds are broken. 
At the same time, the scaling of the modulus is quite different for attractive and repulsive glasses; the attractive glass/gel is much stiffer 
as noted above. 

Thus two states of equal viscosity, close to
the attractive and the repulsive branches of the arrest line, can have quite different relaxation times. 
It would be interesting to measure the dynamics, in the region where the
two branches meet,
along contours of equal zero-shear viscosity, as was done (without shear)
for contours of equal diffusivity in 
recent simulations \cite{Sciortino}.
Because
the yield strain is much smaller 
for a bonded glass than for a caged one, the
nonlinear rheology should vary strongly along these
contours. 

A simple `generalized non-linear Maxwell model'
\cite{fuchsrest} summarizes the pertinent 
behaviour for both AHSS and ISHSM cases:
\begin{equation}\label{maxwell} 
\eta(\dot\gamma) = \eta(\infty) + \left( \frac{1}{\eta(0)}
+ \frac{|\dot\gamma|}{\sigma^*}\right)^{-1}.
\end{equation}
Here $\eta(0)$ is the zero shear viscosity;
$\eta(\infty)$ is a (small) limiting viscosity at very high shear rates (presumably
set by hydrodynamic interactions, ignored here); and $\sigma^*$ is a characteristic
stress scale of order the yield stress 
discontinuity $\sigma^+$ at the onset of arrest.
The above model gives near-plastic yield in arrested systems ($\eta(0) = \infty$) but can also describe a highly viscous but shear-thinning fluid 
phase \cite{fuchsrest}\if{\footnote{A slight improvement in accuracy might be obtained
by replacing $\eta(0)$ with $\eta(0)-\eta(\infty)$ on the right hand
side, but near the transition the effect is minor}}\fi.
In such a phase, close to a repulsive glass transition, the nonlinear viscosity $\eta(\dot\gamma)$ decreases significantly from its
zero-shear value $\eta(0)$ once $\eta(0) \dot \gamma
a^3/kT$ is of order unity.
In a fluid close to an attraction-driven arrest, 
the decrease is only at much higher shear rates, 
with $\eta(0) \dot \gamma a^3/kT \approx 1/\delta$. 
This is associated with the much larger value of the yield stress  ($\sigma^+\sim 1/\delta$), despite the much smaller value of the yield strain 
($u_y\sim\delta$),
in the nearby arrested phase. 

\label{shear}

\section{Concluding remarks}
\label{conclude}
We have described various developments in the application of MCT ideas
to colloids with short range attractions. This theory is very
 successful at high densities in predicting re-entrant transitions
induced by such attractions \cite{pham-etal:2002}; yet this
re-entrance is caused by density fluctuations whose presence partly
undermines the assumptions of homogeneity on which MCT is based. For
a colloid volume fraction of $\phi =0.4$ we found by simulations that
the MCT predictions for averaged properties remain generally good
despite a surprising degree of dynamic heterogeneity. The latter is
washed out in the averages, but can be probed more closely through the
distribution of mean-squared displacements, and this analysis reveals
a strong partitioning of particles into fast and slow populations. At
much lower volume fractions the assumption of homogeneity becomes clearly wrong, but this can be partially addressed, at least at the level of
qualitative prediction, by carefully combining MCT ideas with those of
phase separation (arrest of one phase in a phase-separated morphology)
and/or irreversible aggregation (applying MCT at the cluster
scale). The effects of short range attractions on the rheology of
colloidal suspensions are potentially quite subtle. The preliminary calculations reported above already show some interesting trends, particularly for the
range-dependence of the yield stress in the attraction-driven glass, and for the shear-thinning behaviour in the nearby fluid phase.

\ack

Financial support is acknolwedged from the Ministerio
de Ciencia y Tecnolog\'{\i}a under project MAT2003-03051-CO3-01 (A. M. P.);
Deutsche Forschungsgemeinschaft under Grant No. Fu3093 (M. F.); and EPSRC
GR/S10377/01.

\end{document}